\documentclass[fleqn,12pt]{article}

\usepackage{authblk}

\usepackage{abstract}

\usepackage{graphicx}

\usepackage{hyperref}

\usepackage{amsmath}

\usepackage{indentfirst}

\usepackage[
    left=1.0in,
    right=1.0in,
    top=1.0in,
    bottom=1.0in
]{geometry}

\begin{document}

\title{Classification of Aortic Shape with Topographical Pair Correlation Functions}

\author[1]{Cooper Bruno}
\author[1]{Tiago Cecchi}
\author[2]{Joseph A. Pugar}
\author[2]{Luka Pocivavsek\thanks{Corresponding author}}
\author[3,4]{Newell Washburn\thanks{Corresponding author}}

\affil[1]{Department of Physics, Carnegie Mellon University, Pittsburgh, PA 15213, USA}
\affil[2]{Department of Surgery, University of Chicago, Chicago, IL 60637, USA}
\affil[3]{Department of Chemistry, Carnegie Mellon University, Pittsburgh, PA 15213, USA}
\affil[4]{Department of Biomedical Engineering, Carnegie Mellon University, Pittsburgh, PA 15213, USA}

\date{} 

\maketitle
\thispagestyle{empty}

\begin{abstract}
Quantitative descriptors convert high-dimensional medical images into low-dimensional features capable of differentiating organ shapes that correlate with injury or disease progression for diagnostic purposes. An important example is aortic dissections, which can be imaged using high-resolution CT scans and for which the shape of the true and false lumens of the aorta has long been used to predict disease state and the potential for positive surgical outcomes (namely thoracic endovascular repair or TEVAR). Here we present a method for calculating the topographical pair correlation function (TPCF), a descriptor of the spatial correlation of point estimates for Gaussian curvature, mean curvature, shape index, and bending ratio constrained to the surface of a meshed image. We used the TPCF as a metric to describe aortic shape and extracted quantitative features from the resulting curves. When the TPCF was parameterized by shape index, the area under the curve of the correlation function contributed to a classification accuracy of 0.95 for disease presence and/or impending TEVAR success. Comparison with single-point statistics suggests that the TPCF provides powerful features for classifying the disease state of aortas and more broadly in capturing structural correlations in anatomical data.
\end{abstract}


\section{Introduction}

\indent Quantitative assessment of morphological differences is essential for the classification of 3D images. The general approach reduces a high-dimensional image to a compact set of descriptors then uses these to calculate similarity metrics, allowing probabilistic assessments of injury and disease states. CT images of the liver, the brain, and the vascular system are routinely used in diagnosing injury and disease, but there remains a need for new methods of featurizing them. \\
\indent Given its criticality in the circulatory system, there is extensive literature on structural assessments of the aortic arch that are used in planning surgical procedures to treat aneurysms or dissections. Using 3D multiplanar reconstruction of CT images, traditional approaches to morphometry have focused on arch steepness, take-off angles, and distances between supra-aortic branches \cite{demertzis2010aortic}. Aortic arches are commonly classified into types I, II, and III \cite{Demertzis2010} based on the distances between major branches, which are known to correlate with patient age as well as a diversity of other geometric features, making them a useful predictor for planning surgical aneurysm repair \cite{schumacher1997morphometry}. However, there remains a need for more sophisticated metrics capable of guiding clinical decision-making.\\
\indent Principal component analysis (PCA) is a powerful computational tool for analyzing aortic shape \cite{casciaro2014_modes_variation}. In early applications of PCA, images from 500 asymptomatic patients were segmented into sequentially inscribed circles resulting in 12 geometric parameters that were the basis for PCA, and three components were identified that accounted for ca. 80\% of the variability. This has become the basis for statistical shape analysis (SSA) \cite{Bruse2016}, a broader methodology that connects shape with function \cite{Kiema2022, williams2022shape_hemodynamic} and has been advanced through machine learning approaches \cite{liang2017shape_risk_AsAA} responsible for risk classification accuracies of greater than 95\%. However, these approaches to featurization lack interpretability which makes it difficult to understand the predictions and connect them with other biomechanical models. While maps of the displacement fields allow mode decomposition to extend shape analysis to biomechanics \cite{cosentino2020shape_ATAA}, this methodology still makes interpretation difficult. \\
\indent Point distributions of the curvature have been used by the authors to diagnose aortic disease state \cite{Khabaz2024} as a physics-based featurization method. The fluctuation of Gaussian curvature $\delta K$, calculated as the standard deviation of the point distribution, is a powerful descriptor for distinguishing between healthy and unhealthy aortic shapes --- significantly more so than aortic volume, which increases monotonically during growth and development and thus is insensitive to the formation of structural abnormalities. However, $\delta K$ is insensitive to the spatial correlations between these structural aberrations and thus valuable information is lost during featurization. \\
\indent Continuing with physics-based featurization in this work, we develop a novel method from the pair correlation function. In condensed matter physics, the pair correlation function (PCF) has been used to characterize the structure of disordered systems, ranging from simple liquids to liquid crystals to glasses \cite{hansen2013simple}. The density-density PCF provides a method for calculating important structural parameters, such as the structure of solvation shells in liquids or the spatial fluctuations during phase transitions \cite{chaikin1995pcmp}. It does so by calculating a radial distribution function averaged over each particle, providing the probability of finding an atom at a certain distance given a particle at the origin of the coordinate system. Thus, it is a low-dimensional representation of the spatial connectivity in a complex material that is intermediate between the full list of particle coordinates, which is unwieldy for making predictions, and a single macroscopic scalar, such as the density, which can be incapable of differentiating between distinct, microscopic states.\\
\indent We present the application of a topographical pair correlation function (TPCF) for characterizing the shape of healthy and diseased aortas from CT imaging. In contrast to the traditional PCF calculated in 3D, we have developed the TPCF to be calculated on the surface of a 3D object. Both synthetic shapes – artificial meshes of spheres and cylinders with controlled perturbations – as well as CT scans of healthy and diseased aortas were used, with the synthetic shapes testing and validating the methodology before it was applied to aortic images. Based on point estimates for the Gaussian curvature, mean curvature, shape index \cite{Koenderink1992}, and bending ratio, we demonstrate that the curvature-parameterized TPCF is sensitive to structural differences associated with disease progression and could be a powerful tool in performing similarity analyses of complex 3D images of tissues and organs.

\section{Methodology}

\subsection{Coherence Tomography}
Imaging and clinical outcomes data was acquired from a retrospective single-center aorta registry at the University of Chicago Medical Center (UCMC). The imaging data was acquired through computed tomography angiography (CTA) scans obtained from 2009 to 2020. Subsequent analysis used the scans from 37 non-pathological and 38 pathological patients. Three unique outcome labels are organized by the anatomy segmented from the scan being normal (label 0); non-pathological anatomy of a non-aortic related trauma patient, or the presence (label 2) or absence (label 1) of secondary surgical intervention more than 30 days following the initial thoracic endovascular aortic repair (TEVAR). The decisions to intervene in the case of secondary surgery were made by the primary surgeon. The CTA scans were obtained from the Human Imaging Research Office (HIRO) at the UCMC in accordance with institutional review board approval (IRB \# 20-0653 and IRB \# 21-0299).

\subsection{Meshing and Curvature Calculation}
\subsubsection{Aortic Meshes}
\indent Image processing was performed from raw DICOM files extracted from the CTA data. Segmentation masks (isosurfaces) were created for each aorta from the DICOM files using the Simpleware ScanIP software package (Synopsys, Sunnyvale, CA). After segmentation, recursive Gaussian filtering was applied to smooth each surface to reduce noise while preserving the innate geometry of the segmentation. Subsequently, the built-in FE Free meshing algorithm was used to generate triangular surface meshes which discretized the surfaces. Two approaches were taken while meshing the population of aortic surfaces: enforcing a constant number of elements (number of triangles) or a constant element density (size of triangles). Surface area spanned an order of magnitude (100-1000 cm2) within the dataset, and the impact of each meshing strategy is addressed in the discussion. For each meshed geometry, the Rusinkiewicz algorithm \cite{Rusinkiewicz2004} was used to calculate discrete point principal curvatures at each mesh vertex using a weighted average of shape operators of neighboring elements. The principal curvatures in question are the eigenvalues of the resulting shape operators, and they are the maximum and minimum values of the normal curvature at a given vertex \cite{Rusinkiewicz2004}. The per-vertex Gaussian curvature was calculated as the product of the two principal curvatures,

\begin{equation}
    \kappa_G = \kappa_1\kappa_2.
\end{equation}

Similarly, the per-vertex mean curvature, shape index, and bending ratio were calculated as

\begin{equation}
    H = \frac{1}{2}(\kappa_1 +\kappa_2),
\end{equation}

\begin{equation}
    S = \frac{2}{\pi} \arctan(\frac{\kappa_1 + \kappa_2}{\kappa_1 - \kappa_2}),
\end{equation}

and

\begin{equation}
    R = \frac{\kappa_2}{\kappa_1}.
\end{equation}

respectively. As a result, each aorta was prepared for analysis as an implicit surface with per-vertex discrete curvature metrics.\\

\begin{figure}
    \centering
    \includegraphics[width=0.75\linewidth]{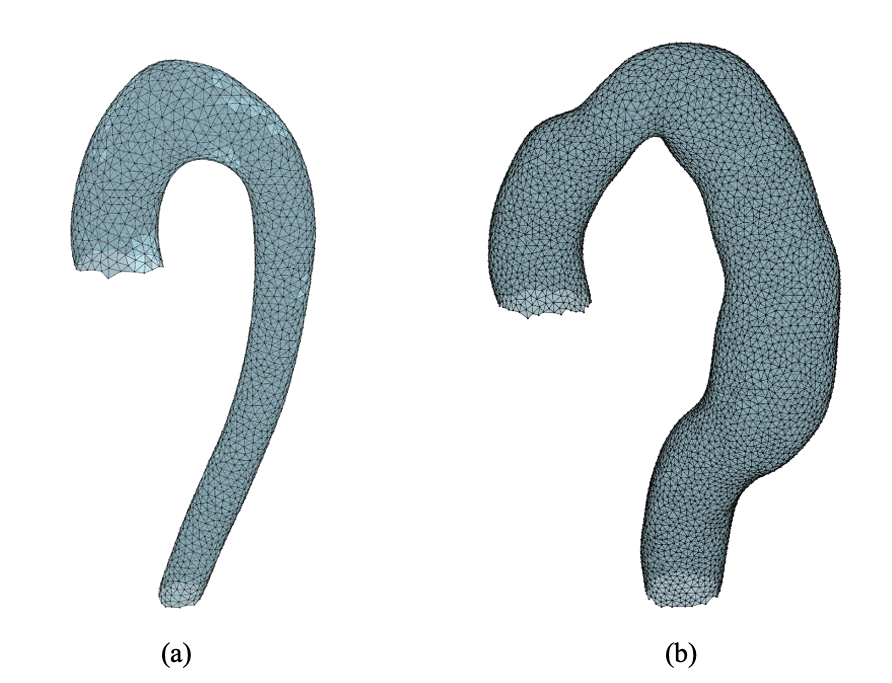}
    \caption{The meshes of two aortas: one healthy (a) and for which no surgical intervention was required, and one unhealthy (b) for which a second surgical intervention more than 30 days following the initial TEVAR was required. }
    \label{fig:placeholder}
\end{figure}

\indent Because diameter has been shown to be a good indicator of the state of an aorta [REF] \cite{Davies2002, Goldfinger2014, Isselbacher2022}, we aim to disregard the physical size of the aortas altogether. Instead, analysis is performed solely based on their shape, and examine the quantities mentioned above parameterized only by distance along the mesh measured in the number of edges. Thus, the methods with which the aortas are meshed prove crucial to our analysis. First, each aorta within the dataset was meshed using a lab-frame approach in which the size of each element was constrained to a preset physical area (a mesh of constant element density). For constant element density meshes, the output signal from the PCF algorithm should indeed convolute the size of the aorta itself, with meshes of larger aortas containing more elements. Therefore, the extrinsic meshes – with a fixed number of elements – are used in the analysis as a baseline for algorithm output: an extension of the ideal shape calibration. To remove mesh density as a confounding variable in the PCF analysis, each aorta was further re-meshed using a quadric error metric decimation algorithm \cite{Garland1997} from open-source Python libraries. The aortas were re-meshed to all contain the exact same number of triangular elements (5000 elements per aorta), regardless of size and thus results in vertices that are relative distances away from one another as size varies.

\subsubsection{Synthetic Shapes}
Prior to the classification of aortic disease state, we validated the use of the TPCF by testing it against expected results for synthetic shapes, including spheres and cylinders. The meshes were created using Trimesh, an open-source Python library. The shapes were spheres and cylinders, which included “idealized” versions as well as variations with varying levels of perturbations to their surfaces. Throughout our analysis, the aorta scans were treated as bent and deformed versions of the synthetic cylinders to explain similarity in the TPCF signal.

\subsection{Topographical Pair Correlation Functions}
\subsubsection{Equation and Notation}

The pair correlation function, or PCF, is a construct that is used throughout condensed matter and biophysics to capture the spatial decay of short-range order in disordered systems. Traditionally represented as a density-density PCF, it is calculated as an average over all particles [replace], counting particles and averaging over volume. \\
\indent We adapt a discrete version as shown in Equation 5.
\begin{equation}
    g(r) = \frac{1}{\sum\limits_k^N{p_k^2}} \sum\limits_{i}^N\left[\frac{1}{n(r,i)}\sum\limits_{j}^{n(r,i)}p_ip_j\right]
\end{equation}

The TPFC, $g(r)$, is a function of distance $r$ measured in edges along the mesh. The number of vertices to which $r$ is the minimum number of edges from vertex $i$ is denoted by $n(r,i)$. All vertices that are reached in a fewer number of edges are neglected in the calculation. Here, $N$ corresponds to the total number of vertices in the mesh, and $p_i$ is the value of a predetermined curvature metric ($\kappa_G$, $H$, $S$, $R$) at vertex $i$. \\
\indent Because the goal is to develop features related to shape and not size, the distance metric in the TPCF of the mesh will remain parameterized by the number of edges and not by a real-space distance. This guided the choice of meshing.

\subsubsection{Featurization and Classification}
\indent Classification of the cohorts of healthy and pathological aortas is done using support vector machines, and a decision boundary is drawn to differentiate the regime in which aortas tend to be healthy from that in which they are typically unhealthy. Across the 38 pathological aortas, 37 healthy aortas, and parameterization by the four distinct curvature metrics, the features used in classifying the aortas are: (1) the area under the TPCF curves for each and (2) mean aortic radius, given in mm.

\section{Results and Discussion}
\subsection{Single-Point Statistics}
Attempting classification only with point statistics (using metrics not explored in previous work) is ineffective. As shown in Fig. 2, there are distinct differences in the distributions of point curvatures between healthy and unhealthy for all curvature metrics. However, this information relays nothing about spatial correlation of curvature and reveals nothing directly about the shape of the aortas. Significant overlap of these distributions for healthy and unhealthy further complicates the problem of classifying disease state without considering spatial correlation.

\begin{figure}
    \centering
    \includegraphics[width=0.9\linewidth]{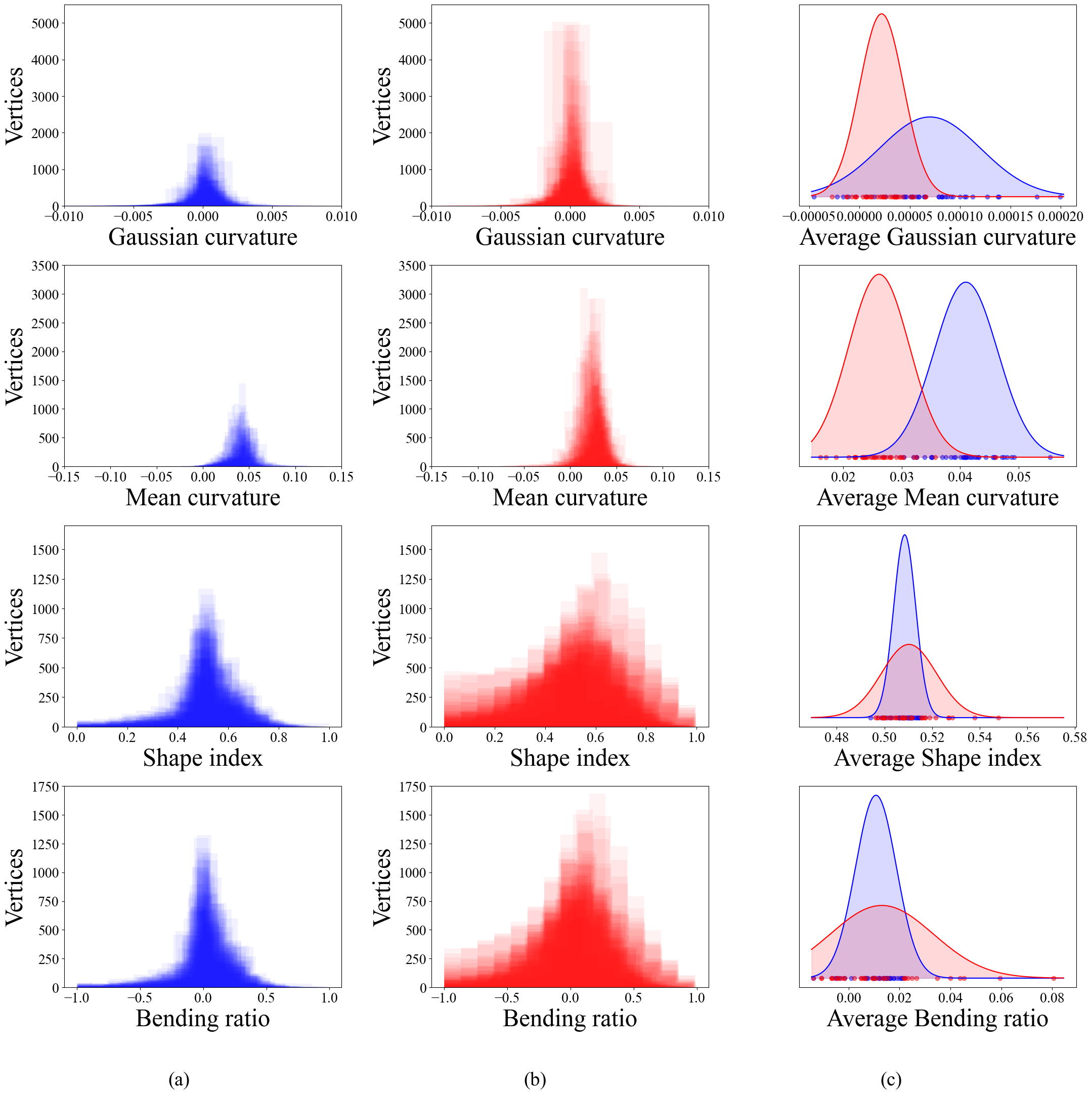}
    \caption{The counts of per-vertex values of all four curvature metrics for (a) healthy and (b) unhealthy aortas. (c) The distributions of average curvature across the healthy (blue) and unhealthy (red) aortas.}
    \label{fig:placeholder}
\end{figure}

\subsection{TPCFs for Synthetic Shapes}
\indent We use here the TPCF to circumvent several challenges posed by analysis of point statistics alone. Through validation with ideal and perturbed synthetic shapes, meshes of spheres and cylinders, we discovered dependence on the spatial frequency and intensity of perturbations along each surface. The TPCF for the perturbed sphere plateaus at a value that depends on the intensity of the perturbations. We also observed a cutoff distance $>$ 30 edges for all shapes, characterized by a sudden decay of the TPCF as a result of fewer accessible vertices, but which depends ultimately upon the density of elements in the mesh. \\

\begin{figure}
    \centering
    \includegraphics[width=0.9\linewidth]{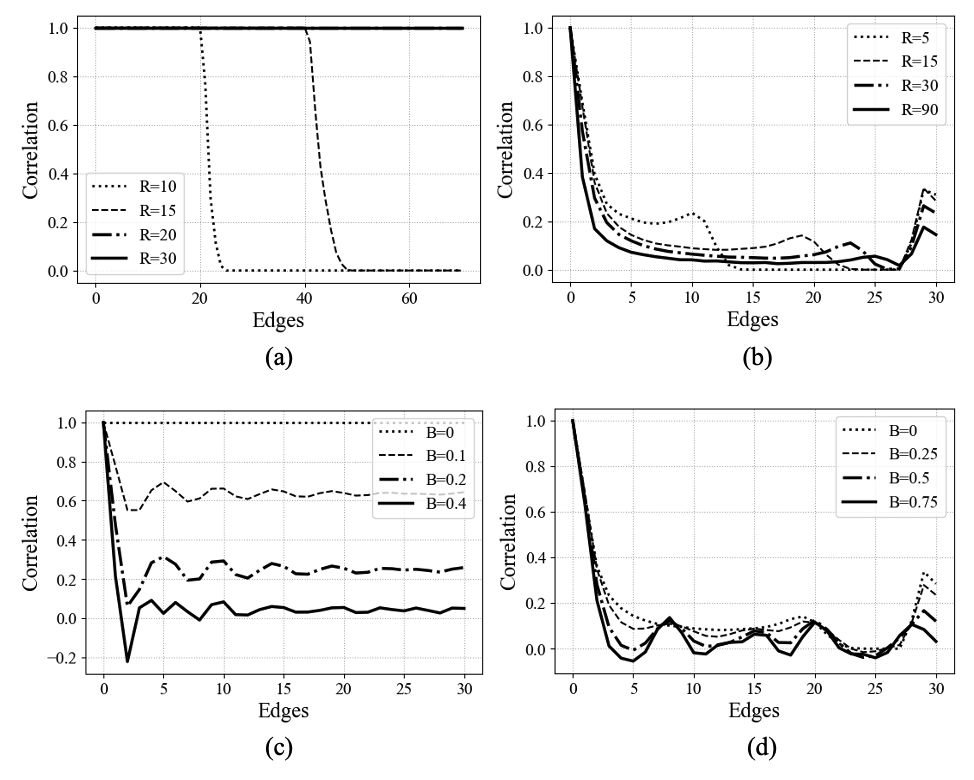}
    \caption{The Gaussian curvature-paramterized TPCFs for (a) ideal spheres and (b) ideal cylinders while varying the radius, and (c) perturbed spheres and (d) perturbed cylinders with fixed radius but increasing intensity of perturbations.}
    \label{fig:placeholder}
\end{figure}

\indent The TPCF for the deformed cylinders is qualitatively very similar to that of the aortas shown later, when parameterized by Gaussian curvature. At short distances, the signal attenuates rapidly, reaches a minimum, and then either begins to increase monotonically again or fluctuates about a near-zero correlation.

\subsection{TPCFs for Thoracic Aortas}
\indent The TPCFs were parameterized by the same four curvature metrics as in Section 3.1 --- Gaussian curvature, mean curvature, shape index, and bending ratio --- and are plotted below for distances from 0 up to 30 edges. We observe similar qualitative behavior between the TPCF parameterized by Gaussian curvature and bending ratio, and similar but more distinctive behavior for shape index and mean curvature. The TPCF signal cleanly separates into two or more bands. \\

\begin{figure}
    \centering
    \includegraphics[width=1 \linewidth]{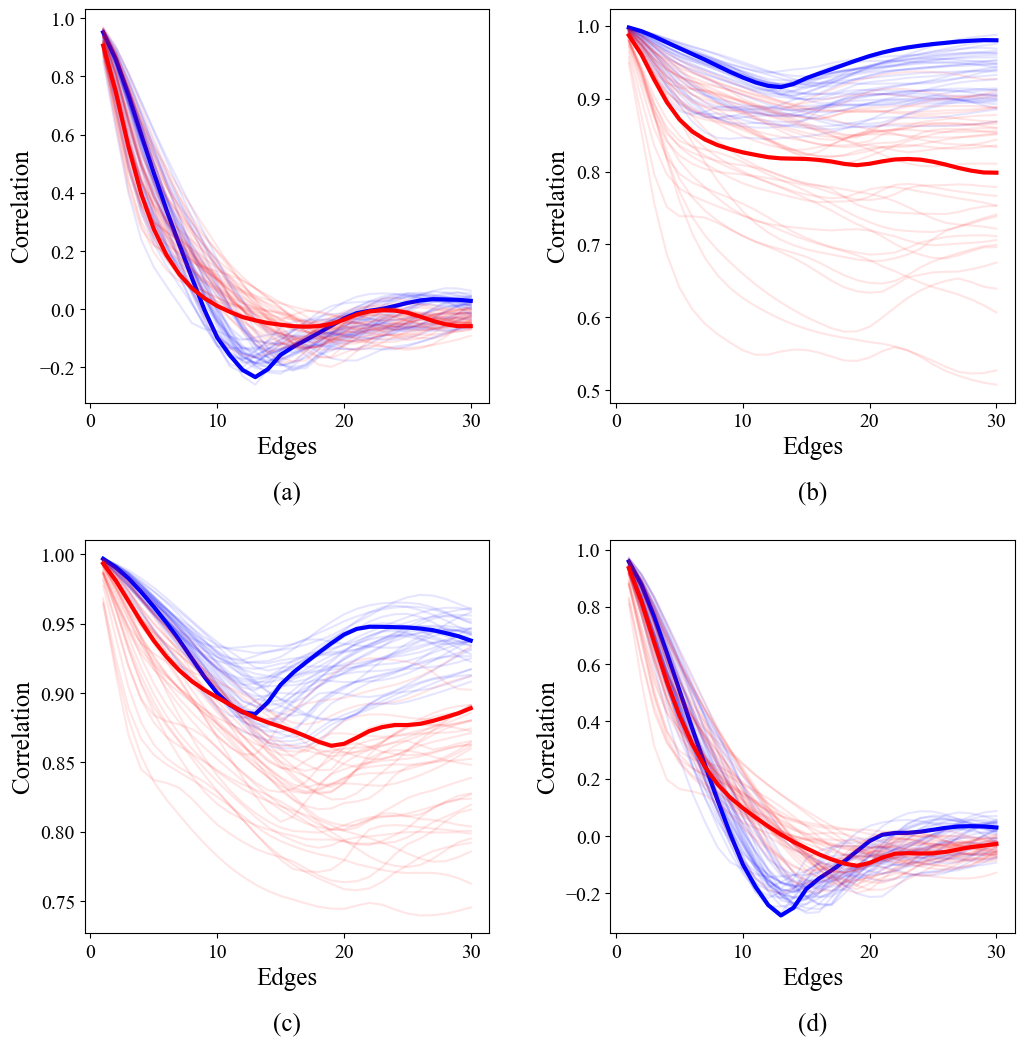}
    \caption{The aortic TPCFs for (a) Gaussian curvature, (b) mean curvature, (c) shape index, and (d) bending ratio. Bolded are the TPCFs for representative healthy (blue) and pathological (red) aortas.}
    \label{fig:placeholder}
\end{figure}

\begin{figure}
    \centering
    \includegraphics[width=1\linewidth]{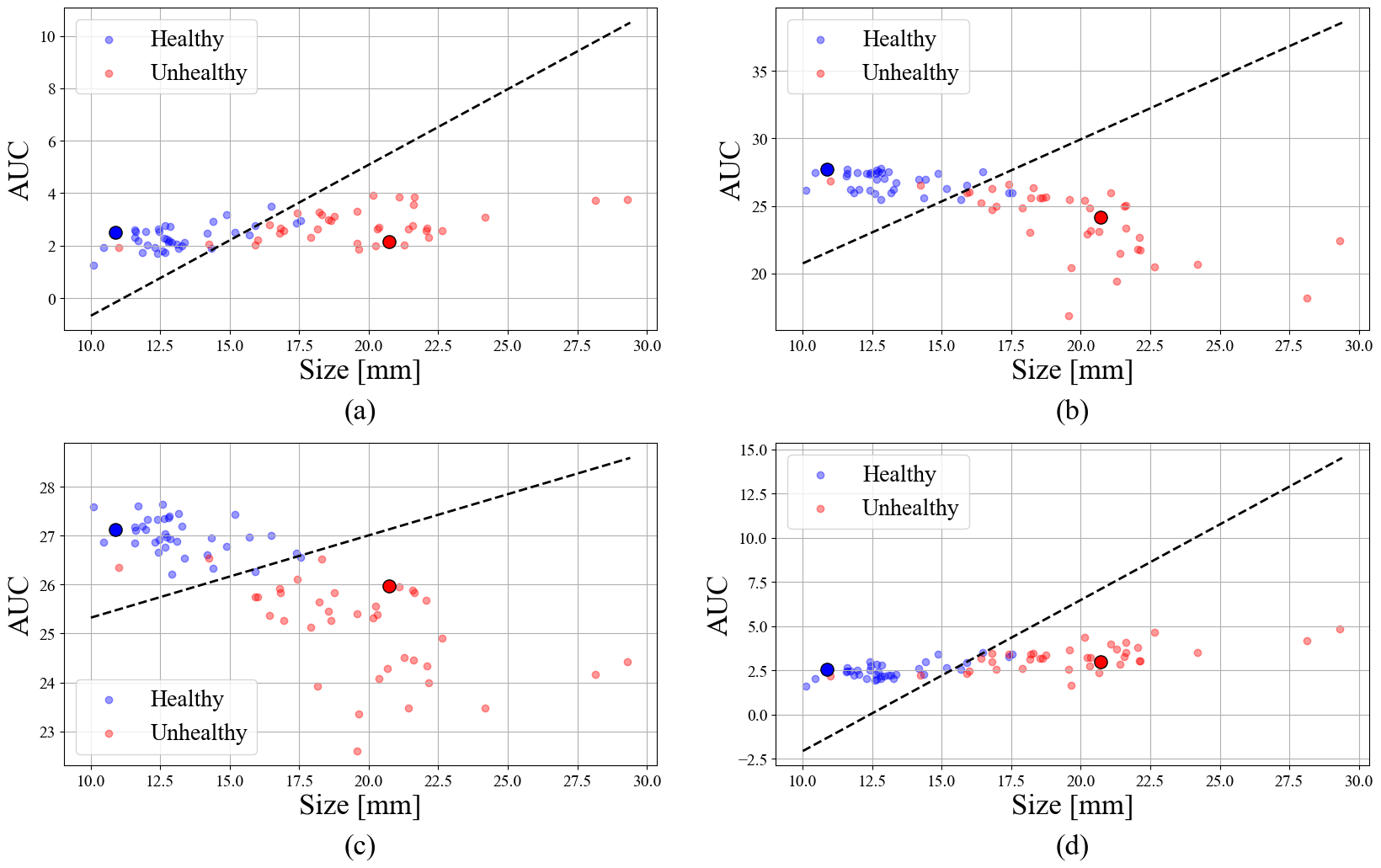}
    \caption{Support Vector Classification performed for TCPF AUC vs. aortic size when the TPCF is parameterized by (a) Gaussian curvature, (b) mean curvature, (c) shape index, and (d) bending ratio. Bolded points represent the same healthy (blue) and pathological (red) aortas as in Fig. 4. }
    \label{fig:placeholder}
\end{figure}

\indent Classification using aortic size and area under the TPCF as the primary features, with an accuracy of 0.95, proved the most effective when the TPCF was parameterized by shape index. Accuracies for Gaussian curvature, mean curvature, and  bending ratio were 0.93, 0.93, and 0.92 respectively.

\section{Conclusion}
\indent The TPCF provides a robust, novel method with which to discriminate between healthy and unhealthy aortas by using morphological features, neglecting the sheer size of an aorta as an indicator of its disease state. Classification accuracies of $\geq$ 0.90 for all curvature metrics indicate that the TPCF is as, if not more, powerful than features based on single-point statistics, such as $\delta K$, when used for classification of disease state. Beyond applications in TEVAR planning, the TPCF is a valuable metric that is sensitive to long-length scale correlations of structural features in a diversity of tissues.

\section{Acknowledgments}

We would like to acknowledge the support of National Institutes of Health grant number R01HL159205 in addition to the Center for Research Informatics, which is funded by the Biological Sciences Division at the University of Chicago with additional funding provided by the Institute for Translational Medicine, CTSA grant number 2U54TR002389-06 from the National Institutes of Health.  Covestro Science Award (NRW).

\bibliographystyle{unsrt}
\bibliography{refs}  

\end{document}